\documentstyle[12pt,aaspp4]{article}

\begin{document}

\title{A SHORT HISTORY OF THE MISSING MASS
AND DARK ENERGY PARADIGMS}
\author{Sidney van den Bergh}
\affil{
Dominion Astrophysical Observatory \\
National Research Council \\
5071 West Saanich Road \\
Victoria, British Columbia \\
V8X 4M6, Canada \\
e-mail: sidney.vandenbergh@nrc.ca }

\bigskip
\bigskip
\begin{quotation}
``An era can be said to end when its basic 
illusions are exhausted''
\end{quotation}

\hspace{3.0in} Arthur Miller
\bigskip
\bigskip

\begin{abstract}
	In 1900 it was believed that almost 100\% of the mass of the
Universe resided in stars.  Now, in the year 2000, such stars (and cold
gas) are known to account for only $\sim$1\% its mass.  The remaining mass of
the Universe is thought to reside in hot baryons ($\sim$3\%), cold dark matter
($\sim$30\%) and dark energy ($\sim$66\%).  The present paper traces the evolution of
our thinking about the density of the Universe during the Twentieth
Century, with special emphasis on the of the discovery of cold dark
matter.
\end{abstract}
						
\clearpage

\section{INTRODUCTION}
\bigskip
\bigskip
\begin{quotation}
``The discrepancy seems to be real and 
important''
\end{quotation}

\hspace{3.0in} Edwin Hubble (1936, p. 180).
\bigskip

When Newton (1687) introduced the notion of gravity he discussed it in
terms of forces between ``bodies'', i.e. visible baryonic objects.  In his
introduction to the {\it Principia} he states that ``I have no regard in this
place to a medium, if any such there is, that freely pervades the
interstices between the parts of bodies.''  I take this to mean that
Newton specifically wished to exclude any consideration of all-pervading
quintessence{\protect \footnotemark[1]}.  
A quarter of a millennium later Zwicky (1933) 
published the first observations that were to overthrow the reigning
paradigm, according to which all gravitational effects were produced by
visible baryonic matter.  Perhaps surprisingly, Zwicky's paper does not
seem to have had significant impact on astronomers during the first half
of the Twentieth Century. Even as late as 1961 only a single paper at the
{\it Santa Barbara Conference on the Instability of Systems of Galaxies}
(Neyman, Page \& Scott 1961) referenced the Zwicky (1933) paper on dark
matter in rich clusters\footnotemark[2].  This paucity of references
cannot just be attributed to the fact that Zwicky's paper was written in
German and published in a relatively obscure (Helvetica Physica Acta)
journal, because Smith's (1936) article in the Astrophysical Journal was
not mentioned by any of the other conference participants either. It is,
however, of interest to note that Edwin Hubble (1936, pp. 180-181) was
aware of the mass discrepancy problem in the Virgo cluster.  He wrote
``The discrepancy seems to be real and important''.  However, it is not
clear from his writings if he also knew about Zwicky's (1933) discovery
of the missing mass problem in Coma.

\footnotetext[1] {According to some Renaissance philosophers, such as Paracelsus,
quintessence ({\it quinta essentia} $=$ fifth essence) was the thinnest and most 
divine material element surrounding the four Empedoclean elements (air, 
earth, fire, and water).  In a related vein Aristotle described the aether as 
the primary substance distinct from the other four.}

\footnotetext[2] {It is of interest to note that the title of the Santa Barbara
conference was ``{\it In}stability of Systems of Galaxies''.  The fact that the
organizers of this meeting thought of groups of galaxies as expanding
positive energy associations, rather than as negative energy stable
clusters, is attested to by the fact that Ambartzumian's name is used
five times in the printed version of the introduction to this conference.}

Unfortunately the Institute of Scientific Information has not yet scanned
the scientific literature prior to 1945.  Data for more recent years are
collected in Table 1.  I am indebted to Helmut Abt, Sharon Hanna and
Sarah Hill for this information.  The table shows a very low citation
rate for Zwicky's pioneering 1933 paper prior to 1975, when the
importance of missing mass began to dawn on the astronomical community.  
A brute force manual search of the Astrophysical Journal for the period
1934-44 revealed only two self-citations by Zwicky (1937, 1942) and a
citation by Smith (1936) in his Virgo cluster paper to Zwicky (1933).

\begin{table}[t]
\smallskip
\caption{Citations of Zwicky (1933)}
\smallskip
\begin{center}
\begin{tabular}{lccr}
\hline
\hline
Year &  & & No. citations \nl
\hline
1955-59	 &  & &     2 \nl
1960-64	 &  & &     6 \nl
1965-69	 &  & &     5 \nl
1970-74	 &  & &     2 \nl
1975-89	 &  & &    63\tablenotemark{a} \nl
1990-99	 &  & &    71 \nl
\hline
\end{tabular}
\tablenotetext{a}{There is a clustering of eight references that cite the 
wrong page number for  Zwicky's article.  Apparently seven of these authors 
copied the reference from Bahcall (1977), which contains a typographical 
error, without actually reading the original paper.}
\end{center}
\end{table}

\section{COLD DARK MATTER}
\bigskip
\bigskip
\begin{quotation}
 ``It is contrary to reason to say that there is 
 a vacuum or a space in which there is 
 absolutely nothing''
\end{quotation}

\hspace{3.0in} Ren\'{e} Descartes

\bigskip
	
From observations of the radial velocities of eight galaxies in the Coma
cluster Zwicky (1933) obtained an unexpectedly large velocity dispersion
$\sigma = 1019 \pm 360$~km~s$^{-1}$.  [This value is, perhaps fortuitously, almost
identical to the modern value $\sigma = 1038 \pm 60$~km~s$^{-1}$ (Colless \& Dunn
1996).] Application of the virial theorem to these data yields [using a
modern distance to Coma] a mass-to-light ratio of $\sim$50 (in solar units).  
This value is an order of magnitude larger than that expected from the
{\it stellar} populations in Coma galaxies.  A similar conclusion was
subsequently reached by Smith (1936) from the radial velocities of 32
members of the Virgo cluster.  In commenting on these results Zwicky
(1957) wrote:  ``It is not certain how these startling results must
ultimately be interpreted.''  Perhaps surprisingly, few astronomers paid
much attention to this alarming result.  When Kahn \& Woltjer (1959)
determined the mass of the Local Group from a timing argument, based on
the observation that M 31 is presently approaching the Galaxy, they did
not reference (and were presumably unaware) of the fact that Zwicky and
Smith had obtained similarly high cluster masses two decades earlier.  
Ambartzumian (1961) tried to explain away the large observed velocity
dispersions in groups and clusters by assuming that clusters of galaxies
(like expanding stellar associations) were unstable, so that the virial
theorem does not apply. However, van den Bergh (1962) pointed out that
this argument must be incorrect because such a large fraction of all
early-type galaxies are presently still members of rich clusters.  This
could not be the case if such clusters were unstable with a short
time-scale.  The situation, as it appeared at the time of the Santa
Barbara conference, was summarized as follows by van den Bergh (1961):  
``(1) The masses of cluster galaxies are too low to prevent such clusters
from expanding rapidly.  (2) The assumption that clusters of galaxies are
expanding rapidly leads to predictions which appear to be in conflict
with observation.''  Einasto, Kaasik \& Saar (1974) first pointed out that
the hot gas (which had been discovered from its X-radiation), does not
have a large enough total mass to stabilize such clusters. This hot gas
does, however, have a larger mass than that which is present in the
stellar populations of cluster galaxies.  Previously, van den Bergh
(1961) had pointed out that the volumes of typical supercompact clusters
[such as Stephan's Quintet] are almost $10^4$ times smaller than those of
normal clusters so that it was difficult to see how they could contain a
sufficient amount of intergalactic material to account for the high
mass-to-light ratios which appear to be indicated by the observations.

A second line of evidence for the existence of large amounts of invisible
mass was developed by Page (1952, 1960), who found that pairs of
elliptical galaxies had a mass-to-light ratio of $66 \pm 14$ (in solar
units).  This showed that such binaries must have massive envelopes, or
be embedded in a massive common envelope.

	The first evidence for the presence of significant amounts of
dark matter associated with an individual galaxy was obtained by Babcock
(1939), who found that the outer regions of M 31 were rotating with an
unexpectedly high velocity.  He concluded that either (1) the outer
region of the Andromeda galaxy has a high mass-to-light ratio, or (2) its
light suffers from unexpectedly large dust absorption.  The latter
explanation could not be applied to the dust-free edge-on S0 galaxy NGC
3115 which was subsequently studied by Oort (1940).  From photometry and
spectroscopy he found that ``the distribution of mass in this object
appears to bear almost no relation to that of light.''  Oort interpreted
this results in terms of a stellar mass distribution that was heavily
weighted towards very faint M-type dwarfs in the outer part of NGC 3115.  
Subsequently Babcock's rotation curve of M 31, and that of Rubin \& Ford
(1970), were extended to even larger radii using 21-cm observations
(Roberts \& Whitehurst 1975) that reached out to a galactocentric distance
of $\sim$30 kpc.  From these observations Roberts \& Whitehurst concluded that
the mass-to-light ratio in the outer regions of M 31 had to be 
$\gtrsim200$.\footnotemark[1]
Following in Babcock's footsteps these authors again assumed that
this high mass-to-light ratio was due to the presence of vast numbers of
dim but massive dM5 stars.

\footnotetext[1]{ Roberts 
(1999, private communication) recalled that his result ``was, at
best, received with skepticism in many colloquia and meeting 
presentations.''}

	Evidence for missing mass in our own Milky Way system was
provided (Finzi 1963) by the fact that the Galactic mass derived from the
motions of distant globular cluster was $\sim$3 times greater than that
obtained from the rotation of the inner disk of the Galaxy.

	The feelings of most astronomers in the early 1960s about the
missing mass problem are, as I remember them, best summarized by de
Vaucouleurs (1960) who wrote:

\begin{description}
\item[``1.] The formation and evolution of groups and clusters of galaxies appear
to be governed by factors besides classical gravitational interaction. 
\item[~\vspace{0.15in}2.] Except possibly for some large, highly condensed clusters of the Coma
type, most groups and loose clusters of galaxies, especially those rich
in spirals, are apparently unstable and may evaporate with lifetimes of a
few billion years, as predicted by Ambartzumian. 
\item[~\vspace{0.15in}3.] If so, cluster masses
derived by application of the virial theorem are illusory. {\bf ''}
\end{description}

	In retrospect it appears that the acceptance of a dark matter
component to the universe was delayed by a decade or so as a result of
the enormously influential paper of Schwarzschild (1954).  Taking direct
aim at Oort (1940), he concluded that ``The observations now available
permit the assumption that in any one galaxy the mass distribution and
the luminosity distribution are identical.  On the other hand the present
observations are not accurate enough to prove this assumption.''  What led
Schwarzschild to this fateful conclusion?  For M 31 the dispersion in
available velocity measurements beyond 80$^{\prime}$ (18 kpc) was so large that a
downturn in the rotation curve of the Andromeda galaxy could not be
excluded.  In the case of M 33, for which the observations only extended
to 30$^{\prime}$ (7 kpc), Schwarzschild concluded that ``the present velocity
observations in M 33 do not disagree with the assumption of identical
mass and light distribution.''  Finally Schwarzschild stated that ``This
bewilderingly high value for the mass-luminosity ratio [in Coma] must be
considered as very uncertain since the mass and particularly the
luminosity of the Coma cluster are still poorly determined.''  In this
connection it is of interest to recall that Zwicky (1937) had, in another
remarkably prescient paper, suggested that gravitational lensing could be
used to determine the true total masses of galaxies and clusters.  He
wrote that ``The observation of such gravitational lens effects promises
to furnish us with the simplest and most accurate determination of
nebular masses.''

	The majority of astronomers did not become convinced of the need
for dark matter in galaxies until the work on the stability of galactic
disks by Ostriker \& Peebles (1973), and the paper in which Ostriker,
Peebles \& Yahil (1974) showed ``that the mass of spiral galaxies increases
almost linearly with radius to nearly 1 Mpc.''  Almost simultaneously
Einasto, Kaasik \& Saar (1974) had also concluded that dynamical evidence
required galaxies to be surrounded by massive ``coronae''.  The modern
interpretation of this missing mass in terms of ``cold dark matter'' (CDM)
is due to Blumenthal et al. (1984).  The term cold dark matter refers to
the fact that CDM presumably consists of slowly moving particles.  The
theory of CDM has so far served us well, although there is some concern
about the fact that it predicts halos (Navarro \& Steinmetz 2000) that are
more centrally concentrated than those that are actually observed.  
Furthermore, the hypothesized CDM appears to have a lumpier structure
than is actually seen.  An additional source of worry is that galactic
disks are smaller than they are predicted to be.  Hogan (2000) has
suggested that endowing cosmic dark matter with a small primordial
velocity dispersion might enable one to preserve the predictions of CDM
on large scales, and improve agreement with observed halo structure.  
Alternatively, Binney, Gerhard \& Silk (2000) have proposed that the
discrepancies between theory and observation may be resolved by taking
into account the effects of massive outflow winds.  Such winds will both
homogenize and smooth the inner halo of a galaxy, and expand it by
absorbing energy and momentum from the ejected material.  

\clearpage

\section{ DARK ENERGY}

\bigskip
\bigskip
\begin{quotation}
``I could be worse employed, than as a 
 watcher of the void''
\end{quotation}

\hspace{3.0in} Robert Frost
\bigskip

After introducing gravity, Newton had to face the question why the
Universe did not collapse under its own gravitation.  His solution to
this problem (Davies 1984) was to assume that, in an infinite universe,
all directions are equal, so that every region experiences an equal pull
in each direction.  When Einstein faced the same problem a quarter of a
millennium later he introduced a cosmic repulsive force as a positive
cosmological constant [and later regretted this as ``the biggest blunder
of my life'' (quoted in Gamow (1970)] into his gravitational field
equations.  In this connection one is reminded of James Joyce's comment
that ``A man of genius makes no mistakes.  His errors are volitional
and are the portals of discovery.''

Observational evidence that is compatible with the existence of dark
energy is provided by (1) the redshift-distance relation for supernovae
of type Ia (Perlmutter et al. 1998, Garnavich et al. 1998), (2)
anisotropies in the cosmic microwave background radiation (Melchiorri et
al. 1999) and (3) gravitational lensing (Mellier 1999).  Presently
available observational constraints suggest (De Bernardis et al. 2000)
that
\begin{equation}
\Omega_{\rm m} + \Omega_{\Lambda} = 1.00 \pm 0.12\ (95\% {\rm\ confidence}), 
\end{equation}
in which $\Omega_{\rm m}$ is the is
the total matter density expressed as a fraction of the 
closure density
of the universe, and
\begin{equation}
\Omega_{\Lambda} \equiv \Lambda/(3 H_0 ^ 2).				
\end{equation}
The data in Eqn. (1) are consistent with the esthetically pleasing
hypothesis that we live in a ``flat'' universe in which
\begin{equation}
\Omega_{\rm m} + \Omega_{\Lambda} = 1.00 .
\end{equation}
With $\Omega\ ({\rm\ cold\ dark\ matter}) = 0.30 \pm 0.10$ and 
$\Omega\ ({\rm baryons}) = 0.04 \pm 0.01$
(Wang et al. 2000), it then follows that 
$\Omega\ ({\rm dark\ energy}) = 0.66 \pm 0.11$;
i.e. 2/3 of the mass of the universe is present in the form of ``dark
energy''.  This energy might be due to a cosmological constant, which is
independent of position and time.  Alternatively it could be in the form
of ``quintessence'' (Caldwell, Dave \& Steinhardt 1998), which is time
dependent and spatially inhomogeneous.  Quintessence clusters
gravitationally on large length scales, but (like a gravitational
constant) remains smooth on small{\protect \footnotemark[1]} 
scales.  Quintessence can
therefore only be felt through its effects on the large-scale dynamics of
the Universe.  Looking toward the future Turner (2000) writes:  ``In
determining the nature of dark energy, I believe that telescopes and not
accelerators will play the leading role''.

\footnotetext[1]{In the present context ``small''  
refers to size-scales smaller than, or
equal to, the dimensions of clusters of galaxies (Ma et al. 1999).}

	It is truly remarkable that the fractions of the mass of the
Universe that are contributed by baryons (4\%), by cold dark matter 
($\sim$30\%) and by dark energy ($\sim$66\%) are all within a factor of only about an
order of magnitude of each other.  In commenting on this strange
coincidence Carroll (2000) wrote:  ``This scenario staggers under the
burden of its unnaturalness, but nevertheless crosses the finish line
well ahead of any competitors by agreeing so well with the data.''

\begin{table}[t]
\smallskip
\caption{Balance Sheet of the Universe}
\smallskip
\begin{center}
\begin{tabular}{lccr}
\hline
\hline
Component & & & Mass fraction \nl
\hline
Stars and cold gas	&  & &  0.01 \nl
Hot gas			&  & &  \underline{0.03} \nl
%			&  & &  ------  \nl
Total baryons		&  & &  0.04 \nl
Cold dark matter        &  & & $\sim$0.30 \nl
Dark energy		&  & & $\sim$0.66 \nl
\hline
\end{tabular}
\end{center}
\end{table}

\section{AFTERTHOUGHTS}
	During the Twentieth Century astronomers have discovered that 
the Universe contains  $\sim$100 times more mass than had previously been  
suspected; most of it in forms that are very difficult to observe 
(see Table 2).  It is slightly 
disconcerting that perhaps only 0.001\% of humanity is presently aware of
the enormous paradigm shift towards a universe in which 99\% is in
{\it in}visible form.  One of the reasons for this is, no doubt, that we live
on Earth, which has a mean density that is almost 30 orders of magnitude
greater than the mean density of the universe.  In other words, we live
in an atypical environment in which cold baryonic matter dominates over
the dark energy of the vacuum by an overwhelming factor.  In some ways
the revolutionary discovery that $\sim$99\% of the mass of the universe is in
invisible form was similar to the quantum revolution of the 1920s, of
which Sam Treiman (1999, p. 18) wrote:  ``There was no immediate commotion
in the streets.  Only a small band of scientists were participating in or
paying close attention to these developments.''

\begin{acknowledgements}
	It is a pleasure to thank Dick Bond, Don Osterbrock, Jim Peebles,
Mort Roberts and Simon White for sharing some of their historical
recollections with me.  I am also indebted to Helmut Abt, to our
librarians Eric LeBlanc and Sharon Hanna, and to NRC librarian Sarah
Hill, for bibliographical assistance.
\end{acknowledgements}

\end{document}